\documentclass[prl,twocolumn,a4paper,superscriptaddress]{revtex4}

\usepackage[pdftex]{color,graphicx}
\usepackage{graphicx}
\usepackage{amssymb}
\usepackage[normalem]{ulem}
\usepackage{color}
\usepackage{ulem}

\begin{document}

\title{Stable room-temperature ferromagnetic phase at the FeRh(100) surface }

\newcommand{\lsi}{Laboratoire des Solides Irradi\'es, \'Ecole Polytechnique, CNRS, CEA-DSM-IRAMIS,  Universit\'e Paris-Saclay, F-91128 Palaiseau, France}
\newcommand{\etsf}{European Theoretical Spectroscopy Facility (ETSF)}
\newcommand{\soleil}{Synchrotron SOLEIL, L'Orme des Merisiers, Saint-Aubin, BP 48, F-91192 Gif-sur-Yvette, France}
\newcommand{\ucsd}{Center for Magnetic Recording Research, University of California, San Diego, 9500 Gilman Drive, La Jolla, California 92093-0401, USA}

\author{Federico Pressacco}
\affiliation{\soleil}
\author{Vojt\v{e}ch  Uhl\'i\v{r}}
\affiliation{\ucsd}
\author{Matteo Gatti}
\affiliation{\lsi}\affiliation{\etsf}\affiliation{\soleil}
\author{Azzedine Bendounan}
\affiliation{\soleil}
\author{Eric E. Fullerton}
\affiliation{\ucsd}
\author{Fausto Sirotti}
\affiliation{\soleil}

\date{\today}

\begin{abstract}

Interfaces and low dimensionality are sources of strong modifications of electronic, structural, and magnetic properties of materials.
FeRh alloys are an excellent example because of the first-order phase transition taking place at $\sim$400 K from an antiferromagnetic phase at room temperature to a high temperature ferromagnetic one.
It is accompanied by a resistance change and  volume expansion of about 1\%.  
We have investigated the electronic and magnetic properties of FeRh(100) epitaxially grown on MgO by combining spectroscopies characterized by different probing depths, namely X-ray magnetic circular dichroism and photoelectron spectroscopy.
We thus reveal that the symmetry breaking induced at the Rh-terminated surface stabilizes a surface ferromagnetic layer involving five planes of Fe and Rh atoms in the nominally antiferromagnetic phase at room temperature. First-principles calculations provide a microscopic
description of the structural relaxation and the electron spin-density distribution that fully support the experimental findings.
\end{abstract}

\maketitle

Electronic and magnetic properties are strongly modified by symmetry breaking and reduced dimensionality at surfaces and interfaces. 
Spatial confinement and interface engineering can lead to fundamental
discoveries of new phases and functionalities (e.g. topological insulators
\cite{Hasan2010} or interface phenomena in complex oxides \cite{Coey2013,Zubko2011}) revealing
emerging behavior that is not present or is very different in the bulk \cite{Hwang2012,Jian2013}.
In this regard the  FeRh compound is a promising material  showing a metamagnetic first-order phase transition above room temperature that is of great interest for future  technologies such as heat assisted magnetic random access memories (HA-MRAM) \cite{Bennet,Kande}, magnetic cooling \cite{entropy1,entropy2} and spintronics devices \cite{Matri,Cherifi2014,bordel}.
The transition is a change of the magnetic ordering of the Fe moments from antiferromagnetic (AFM) at room temperature to ferromagnetic (FM) above 380 K, which is  followed by the appearance of a net magnetic moment on the Rh atoms \cite{Fallot1939,ZakharovFeRh,stamm2008}.
This metamagnetic transition is commensurate with an isotropic expansion of the lattice structure \cite{ibarra} and a sizeable variation of the magnetoresistance \cite{Suzuki2011}.   
The complexity of the phenomenon therefore raises fundamental questions on the interplay between the magnetic order, the electronic properties and the atomic structure. 
Thsi has triggered a number of studies and experiments over the last decades with the aim to explore the nature and driving mechanisms of this magnetostructural transition \cite{Kittel1960,Tu1969,Mariager-Pressacco,Bergman2006,Gruner2003,Gu2005,Sandratskii2011}.  

While most of the studies were focused on the behavior of bulk samples, in recent years particular interest has grown around the properties of FeRh in presence of interfaces with substrates \cite{Fan-interface} and overlayers \cite{Ding2008,baldasseroni,Loving2012,Han2013,baldasseroni_jap}, also with the aim to identify their influence on the nucleation of the FM domains associated with the phase transition \citep{Bergman2006,Kim2009,Baldasseroni2012}. 
The identification of an interfacial FM layer was attributed to a combination of the effects of strain, Fe deficiency and chemical diffusion from the overlayer.
On the other hand, only few experiments on the free surface of FeRh are reported in literature\cite{Gray2012,Lee2010}.
In particular, using ultrathin films of FeRh epitaxially grown on a W(100) single crystal, Lee and coworkers  \cite{Lee2010} investigated the spin polarization of the valence band of the Fe-terminated surface. 
Spin polarized angular resolved photoemission experiments concluded that the temperature dependence of the surface magnetic properties is the one expected for the bulk and no evidence of a privileged magnetic state at clean surface has been observed. 
 
The physical properties of a material are strongly dependent by the atomic distribution and relaxation at surfaces and interfaces.
In this Letter, we present a detailed study of the electronic and magnetic properties of a high quality  FeRh epitaxial layer,grown on MgO(100), terminated by a plane of Rh atoms. 
Using synchrotron-radiation spectroscopy techniques with different probing depths we demonstrate the presence of a stable ferromagnetic surface for FeRh at room temperature while the bulk is antiferromagnetic. 
This experimental finding agrees with first-principles calculations which give a detailed description of the atomic, electronic and spin distribution at the material/vacuum interface.

\begin{figure}
  \vskip -0.0cm
    \begin{center}
          \includegraphics[width=\linewidth]{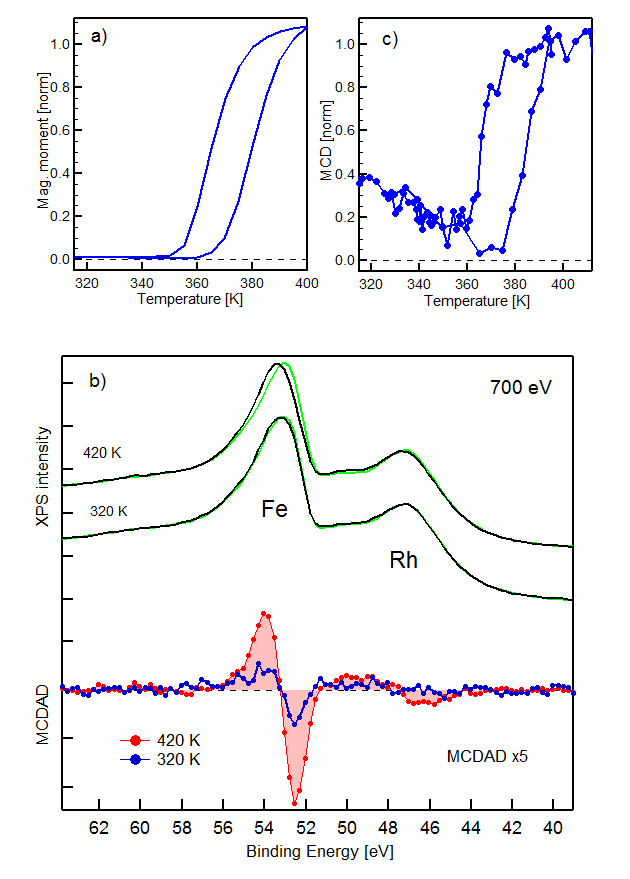}
    \end{center}
    \vskip -0.5cm
 \caption{(a) Temperature dependence of the magnetization measured by vibrating sample magnetometry. 
 (b) Magnetic circular dichroism of FeRh core levels measured by photoelectron spectroscopy at 700 eV excitation energy. 
 The blue curve corresponds to the signal in the ferromagnetic phase at 420K, while the red curve shows the residual magnetism at 320 K.
 (c) The MCDAD contrast extracted from (b) is plotted as a function of temperature in a heating-cooling cycle.
  }
 \label{fig:fig1}
\end{figure}

FeRh(100) films, 50 nm thick, were grown epitaxially on MgO(100) substrates by dc magnetron sputtering using an equiatomic target. The films were grown at 450 °C and post-annealed at 800 °C for 45 minutes..
Temperature-dependent vibrating sample magnetometry (VSM)(see Fig. \ref{fig:fig1}(a)) shows the expected hysteretic behavior of the sample magnetization 
suggesting a homogeneous and ordered B2 phase \cite{maat,back2004,bergman}.
The samples were protected with a 2 nm thick Pt layer grown after cooling down to room temperature.
In the preparation chamber of the TEMPO beamline at the SOLEIL synchrotron radiation source, the capping was removed by low-energy (500 eV) Ar sputtering followed by annealing at 600 K for 30 minutes.
The prepared samples were kept in a vacuum better than $5\times10^{10}$ mbar for spectroscopic investigation. 
Quantitative analysis  of the  Fe-3p and Rh-4p core levels X-ray photoemission spectroscopy  (XPS)  measured at minimum electron inelastic mean free path $\Lambda_{\text{IMFP}}$ demonstrates a Rh-terminated surface (see Section I of \cite{suppmat}).
The Rh-terminated surface could be reproduced on several samples, indicating a stable configuration.
The surface showed very stable electronic and magnetic properties and low reactivity since no oxidation or carbon contamination occurred over several hours.

Circularly polarized soft X-rays from the TEMPO beamline \cite{TEMPO} were used to study the magnetic properties of the FeRh surface as a function of temperature by measuring spectroscopic signals characterized by different probing depths.
In particular, X-ray magnetic circular dichroism (XMCD) \citep{Thole-xmcd,Stohr-saturation} experiments at the Fe $L_{2,3}$ and Rh $M_{2,3}$ edges performed detecting the sample photocurrent have a probing depth of about 3 nm \cite{Abbate1992}. 
In core-level photoemission spectroscopy we measured the magnetic circular dichroism in angular distribution (MCDAD), which is known to be proportional to the sample magnetization. 
In this case the mean free path, and hence the probing depth, can be tuned by selecting the photon energy.
The XPS signal from the Fe 3p core level \citep{Baumgarten} measured with excitation energy of 700 eV has a probing depth of about 1.2 nm \cite{IMFP}.
In all the experiments the FeRh film is magnetized in the horizontal plane and the circularly polarized  photon impinges on the sample with an angle of 42 degree. 
With this geometry, the electron-energy analyser measures photoelectrons along the surface normal.
Because of the constraint induced by photoelectron-spectroscopy experiments, all the measurements are performed at magnetic remanence state, after removing applied magnetic field (400 Oe applied for 300 ms).

In Fig.\ref{fig:fig1}(b) we report the xps spectra of Fe 3p and Rh 4p core levels measured at 420 and 320 K with an excitation energy of 700 eV.
The black and gray curves refers to opposite and parallel  directions of the magnetization with respect to photon helicity.
The difference curves are reported at the bottom of Fig.\ref{fig:fig1}(b).
The red curve is the MCDAD signal extracted from the spectra when the system is in the FM phase.
A residual dichroism is observed at 320 K (blue curve), it corresponds to about 25\% of the one measured for the FM phase.
Similar behavior is observed also for the Rh 4p core level, but the low magnetic signal can not be used for a  quantitative evaluation.
The magnetic signal observed at 320 K is obtained when the sample in the AFM phase, well below the transition temperature. 
We have measured the temperature dependence of the MCDAD contrast by reversing the magnetization after each scan while slowly cycling the sample temperature between 420 and 300 K,  see Fig. \ref{fig:fig1}(c). 
Each point corresponds to the integral of the positive and negative lobes of the MCDAD curve \cite{sirotti}. 
The temperature dependence of the FeRh bulk magnetization is well reproduced: the transition temperature found in XPS are consistent with the VSM measurement made to characterize the system, see Fig.\ref{fig:fig1}(a).  
Starting from the FM phase, in the cooling branch a strong reduction of the signal is observed, but the signal recovers back to 25\% of the maximum. 
In the heating branch a reduction of the signal is observed until 380 K where a rapid increase up to the maximum takes place.
The observed reduction of the signal is consistent with an increase of the disorder of the moments while approaching the transition temperature.
\begin{figure}
  \vskip -0.0cm
    \begin{center}
          \includegraphics[width=\linewidth]{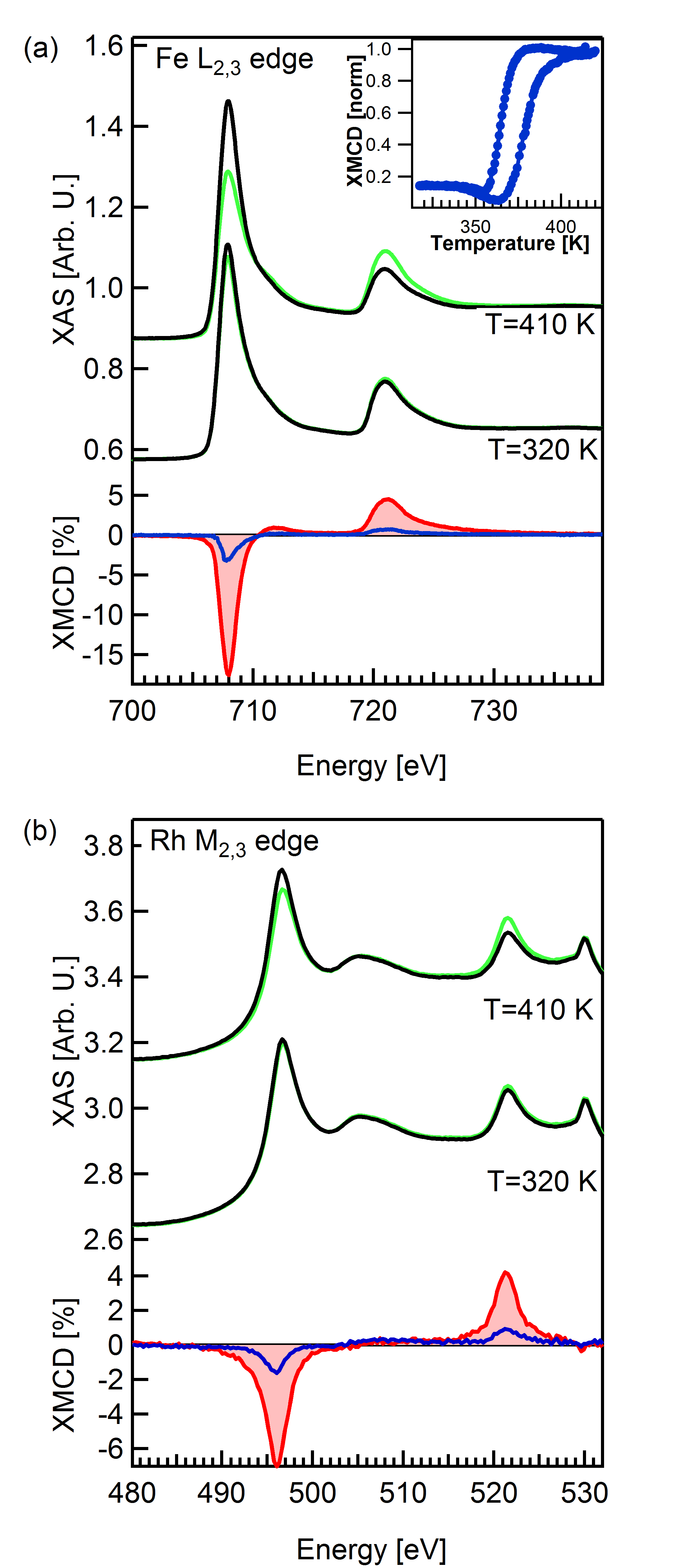}
    \end{center}
    \vskip -0.5cm
 \caption{(a) X-ray absorption (top) and XMCD (bottom) of Fe $L_{2,3}$ edges at 410K, 320K. The dichroism at 320 K (blue curve) is 18 \% of that measured in the FM phase (red filled curve). (Inset) The XMCD dichroism measured at the  Fe $L_{2}$ edge  as a function of temperature in a heating-cooling cycle.   (b)  X-ray absorption (top) and XMCD (bottom) of Rh $M_{2,3}$ edges at 410K, 320 K. For Rh the residual dichroism at 310 K is 23 \% of that in the FM phase.}
 \label{fig:fig2}
\end{figure}

To localize the origin of the magnetic signal measured at 320 K analogous experiments were performed using XMCD from the  X-ray absorption spectroscopy (XAS) at the Fe $L_{2,3}$ and Rh $M_{2,3}$ edges, which is less surface sensitive. 
The XAS spectra measured at 410 and 320 K with opposite magnetization directions are reported in  Fig. \ref{fig:fig2} (a) and (b)  for Fe and Rh respectively. 
The XMCD difference signals, in an enlarged scale, are presented in the bottom. 
By applying the XMCD sum rules to the FM phase spectra, and  by taking into account the incidence angle of the X rays with respect to the magnetization direction we obtain $2.9\, \mu_{B}$ for the Fe magnetic moment.
It can be compared with the value given by {\it ab initio} calculations at 0 K of $3.2\, \mu_{B}$  \cite{suppmat}.
The difference is in good agreement with a 10\% reduction expected for a sample with Curie temperature of 675 K \cite{Kouvel1961}.
Again a magnetic signal is present at room temperature (320 K, blue XMCD curve) as in the case of photoemission, but now the amplitude relative to the signal measured in the FM phase at high temperature (410K, blue XMCD curve) is 18\%.
A magnetic signal equal to 23\% of the remanent FM value is present also for Rh edge at 320 K, higher then the one observed for Fe.  
By fixing the photon energy at the maximum of the XMCD signal and reversing the sample magnetization direction with the same procedure used for the photoemission experiment, we measured the temperature dependence of the magnetic signal. 
The result, which is presented in the inset of Fig. \ref{fig:fig2}(a), reproduces the temperature dependence of the MCDAD contrast in Fig. \ref{fig:fig1}(c).

The measured values of the magnetic dichroism can be fully understood by taking into account the different probing depths of x-ray absorption and photoemission.
The contribution of each atomic plane (Fe or Rh) to the total magnetic signal can be calculated by introducing an exponential weight (see Section II of \cite{suppmat}).
With a decay length of 1.2 nm for the photoemission and 3.4 nm for x-ray absorption [33], the measured magnetic signals can be reproduced assuming an antiferromagnetic bulk and a surface ferromagnetic phase composed by the first 2 Fe and 3 Rh atomic planes.

\begin{figure}
  \vskip -0.0cm
    \begin{center}
          \includegraphics[width=\linewidth]{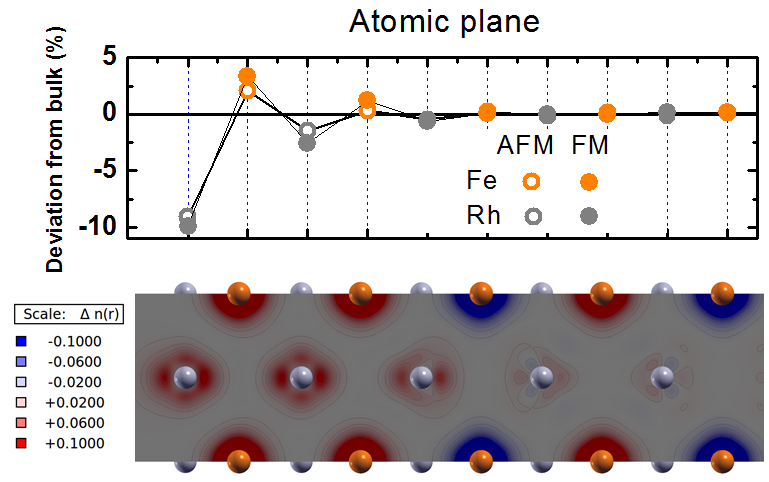}
    \end{center}
    \vskip -0.5cm
 \caption{(Top panel) Calculated atomic relaxations  for the Rh-terminated surface (which is located on the left side of the figure). 
 The percent deviation from the bulk  is represented with solid and empty symbols for the FM and AFM phases, respectively. 
 The positions of the  5 topmost atomic planes are significantly modified. 
 In particular, the Rh planes undergo a compressive displacement while the Fe planes are expanded. 
 (Bottom panel) Map of the calculated spin-density distribution in a red-grey-blue colorscale of an AFM slab with a FM layer at the Rh-terminated surface and an AFM configuration below the third Fe plane (see text). 
 Grey and red spheres identify Rh and Fe atoms aligned with the top panel, respectively. 
 A net spin polarization is clearly observed for the 3  Rh planes  closest to the surface (left side).}
 \label{fig:fig3}
\end{figure}

To gain a microscopic confirmation of this experimental indication, we performed first-principle calculations \cite{suppmat} within density functional theory (DFT) in the local-density approximation (LDA) \cite{Kohn1965}. 
We simulated the FeRh(100) surface by considering a slab consisting of 16 alternating atomic planes of Fe and Rh in the FM and type-II AFM configurations \cite{Moruzzi1992}.
In the latter, Fe atoms are aligned antiferromagnetically within each layer containing two inequivalent atoms per unit cell.
In this slab model one surface is Rh-terminated and the other is Fe-terminated.
The top panel of Fig. \ref{fig:fig3} shows the displacements in the direction perpendicular to the surface of Fe and Rh atoms in the slab model (represented as red and grey circles respectively) as an effect of the structural relaxation from bulk positions (only the topmost 10 planes starting from the Rh surface, on the left of the figure, are shown for better clarity).
We find that while the central planes  remain equidistant as in the bulk, the surface planes undergo a sizeable displacement, confirming that the surface properties in FeRh can be expected to differ from the bulk.
In particular, for both Fe- and Rh-terminated surfaces the most external Rh planes move toward the bulk (the deviations from bulk position are negative, top panel of Fig\ref{fig:fig3}) while the Fe planes move toward the vacuum (the deviations from the bulk are positive).
This occurs independently of the magnetic configuration, as for both AFM and FM qualitatively similar displacements are observed.
As a result of this structural relaxation, while surface interplanar Rh-Rh distances decrease, Fe-Fe distances increase.
Since also in the bulk FM phase Fe-Fe distances are greater than in AFM phase, this result suggests that from the structural point of view  the surface layers relax towards the bulk FM state. 

In the relaxed structures, besides the FM and AFM configurations, we also simulated a mixed magnetic configuration in which the 2 topmost Fe layers in the Rh terminated surface are ferromagnetic and the remaining 6 are antiferromagnetic. 
The different self-consistent solutions at 0 K are found to be within 9 meV/atom energy, suggesting a phase space with several local minima separated by shallow barriers.
Moreover, in all simulations the initial magnetization for Rh atoms was always set different from zero. 
In the fully AFM configuration, both in the bulk and in the slab, Rh local moments at self-consistency always converge to zero. 
This is due to a magnetic frustration induced by hybridization with Fe atoms with opposite moments located at equal distances from Rh atoms \cite{Sandratskii2011,Kudrnovsky2015}. 
On the contrary, in the slab with the mixed magnetic configuration, Rh atoms in the surface layers still preserve a finite magnetic moment, as shown by the bottom panel of Fig. \ref{fig:fig3}, where the magnetization density is depicted (red and blue contours represent positive and negative values, respectively). 
We find that in the 2 topmost surface Rh layers (grey balls) the magnetic moments (1.0\, $\mu_{B}$ each, as in the bulk FM phase) are ferromagnetically aligned. Moreover, also in third Rh layer, which is sandwiched between a FM and an AFM Fe layer (red balls), a finite magnetic moment of 0.6 $\mu_{B}$ still survives. In the other Rh layers instead the magnetic moments go to zero (even though the magnetization density around Rh atoms is not zero, as discussed in \cite{Sandratskii2011}). 
The results of the calculations therefore confirm the existence of the FM surface layer (with the magnetic moments of the bulk FM phase) that has been experimentally discovered.

In conclusion, we have given an experimental evidence that the Rh-terminated surface of FeRh is ferromagnetic at room temperature while the bulk is in the AFM phase.
A simple phenomenological model based on the assumption of different probing depths for XAS and photoemission indicates that  5 atomic planes, 3 of Rh and 2 of Fe, are ferromagnetic. 
First-principle calculations provide a microscopic description of the surface structural relaxation and of the spin-density distribution confirming the stability of the 5 FM planes at the Rh-terminated surface, and 
supporting the interpretation of the experimental data.

These results show how surface and interface relaxation can induce specific electronic and magnetic configurations and explain how several results attributed to chemical bonding with capping layer can be also related to the relaxation process at the interface.
Furthermore the findings suggest that it is possible to tailor a well-confined magnetic thin layer by modifying the atomic arrangement of the sample.
The study of the modifications induced by the atomic interaction with other materials is necessary for a complete understanding of the interface magnetic behavior. 

V.U. and E.E.F. acknowledge support from the U.S. Department of Energy, Office of Science, Office of Basic Energy Sciences under Contract No. DE-SC0003678.

\bibliographystyle{unsrt}
\bibliographystyle{apsrev} 

\bibliography{./FeRh_Surface_Mag_biblio}

\end{document}


\title{SUPPLEMENTARY MATERIAL TO ``Stable room-temperature ferromagnetic phase at the FeRh(100) surface''}

\newcommand{\lsi}{Laboratoire des Solides Irradi\'es, \'Ecole Polytechnique, CNRS, CEA-DSM-IRAMIS,  Universit\'e Paris-Saclay, F-91128 Palaiseau, France}
\newcommand{\etsf}{European Theoretical Spectroscopy Facility (ETSF)}
\newcommand{\soleil}{Synchrotron SOLEIL, L'Orme des Merisiers, Saint-Aubin, BP 48, F-91192 Gif-sur-Yvette, France}
\newcommand{\ucsd}{Center for Magnetic Recording Research, University of California, San Diego, 9500 Gilman Drive, La Jolla, California 92093-0401, USA}

\author{Federico Pressacco}
\affiliation{\soleil}
\author{Vojt\v{e}ch  Uhl\'i\v{r}}
\affiliation{\ucsd}
\author{Matteo Gatti}
\affiliation{\lsi}\affiliation{\etsf}\affiliation{\soleil}
\author{Azzedine Bendounan}
\affiliation{\soleil}
\author{Eric E. Fullerton}
\affiliation{\ucsd}
\author{Fausto Sirotti}
\affiliation{\soleil}

\date{\today}

\date{\today}

\maketitle

\section{Surface Termination}

\begin{figure}
  \vskip -0.0cm
    \begin{center}
          \includegraphics[width=\linewidth]{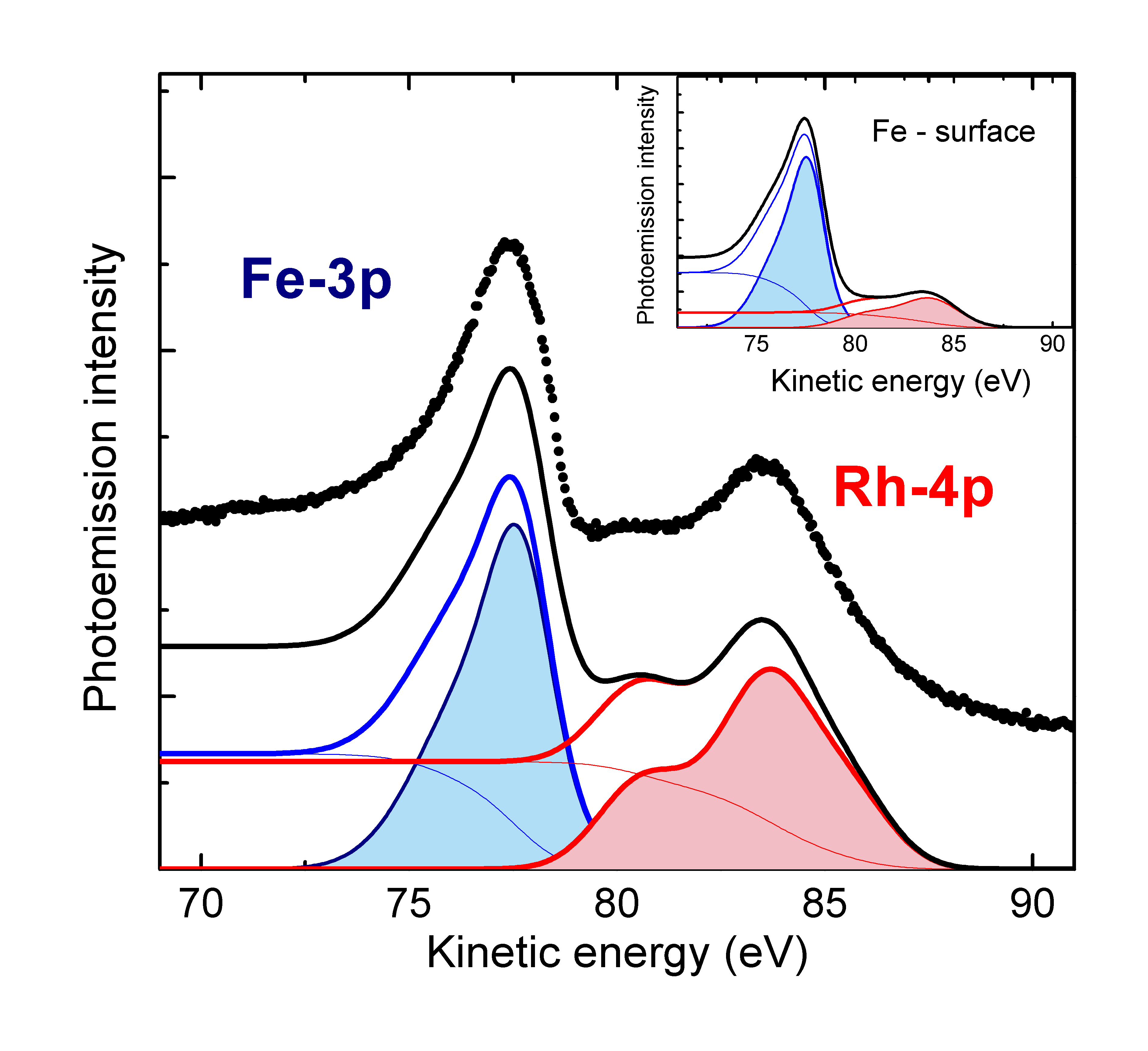}
    \end{center}
    \vskip -0.5cm
 \caption{: Fe 3p and the Rh 4p core levels measured on the Rh-terminated FeRh system (black dots). The black line is the calculated photoemission intensity using Gaussian line shapes and steps functions (thin red and blue lines) obtained by integration to simulate the corresponding secondary electron background. Red and blue filled curves are the background subtracted spectral shapes used to calculate the intensity ratio. (Inset)Simulation of the photoemission spectrum for a Fe-terminated FeRh surface, to be compared with Fig. 3 of \cite{Lee2010}. 
}
 \label{fig:fig1}
\end{figure}

In the direction normal to the surface, the FeRh film looks like a stacked material composed by alternating planes of Fe or Rh.
We determine the atomic arrangement of the sample by comparing the intensities of  Fe 3p and Rh 4p photoemission peaks ( dotted line in Fig.\ref{fig:fig1}.
Since these two peaks have similar binding energies (Fe 3p 52 eV and Rh 4p 46 eV), the photoemitted electrons have similar inelastic mean free path $\Lambda_{\text{IMFP}}$.
When using 130 eV X-ray photons, $\Lambda_{\text{IMFP}}$ is close to the minimum, corresponding to $4.5$  \AA \,  for FeRh \cite{IMFP}.
We use two Gaussian lineshapes with variable spin-orbit splitting, one for each element, to reproduce the spectrum.
More complex configurations \cite{Sirotti,VanderLaan} would not be justified for our quantitative analysis.
The Fe and Rh components of the fit results are presented as filled color curves in Fig.\ref{fig:fig1}.
A background step function (thin lines) was obtained by integrating the primary peak for each element.
The intensity and the width of the Gaussians were optimized to reproduce the measured photoemission intensity (black line).
The fit procedure allows us to calculate the relative intensity of the Rh and Fe peaks.
The ratio between the purple and the blue areas in Fig.\ref{fig:fig1} is 0.92. 
Because of the simple atomic arrangement of the Fe and Rh atoms along the FeRh[100] normal we can model the relative photoemission peak intensity by summing the contributions $I_i$ of each atomic plane at position $z_i$
weighted by an exponential function with  $\Lambda_{\text{IMFP}}$ as decay constant:
\begin{equation}
I_{i}(z_{i},N_{a},E)=\sigma_{i}(E,N_{a}) \exp^{-\frac{z_{i}(N_{a})}{ \Lambda_{\text{IMFP}}}}
\label{eq:equa1}
\end{equation}
.
We have used tabulated  photoionization cross sections \cite{webcrossections} $\sigma_{i}$ for each element $N_{a}=\text{(Fe,Rh)}$, photon energy $E$, 
and calculated electron inelastic  mean free path \cite{IMFP} $ \Lambda_{\text{IMFP}}$. 
We applied this model considering the two possible ideal terminations of the film.
The value obtained for the Rh-terminated surface is 0.88, which is in good agreement with the value extracted from fit parameters of the experimental spectrum in Fig.\ref{fig:fig1}.
The discrepancy with the value extracted from the measurement indicates a slight Rh abundance.  
For a Fe-terminated surface the model instead predicts a ratio of 0.21. 
In the inset of Fig.\ref{fig:fig1} we report the simulated spectrum for a Fe-terminated surface.
The result is in good agreement with the measurement on FeRh grown on W(100) single crystal done by Lee and coworkers \citep{Lee2010}.
\section{Simulation of the magnetic planes moment orientation}
\begin{figure}
  \vskip -0.0cm
    \begin{center}
          \includegraphics[width=\linewidth]{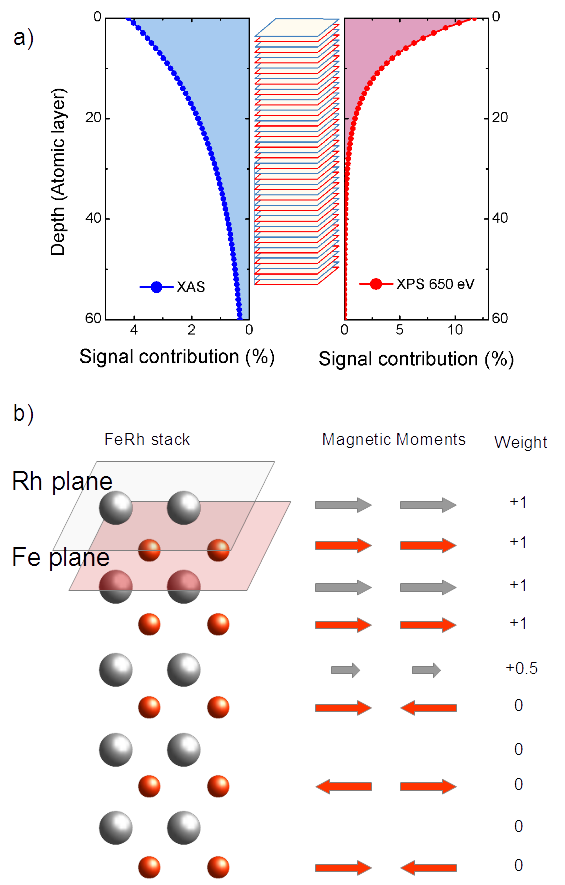}
    \end{center}
    \vskip -0.5cm
 \caption{(a )Contribution to the total signal intensity of each atomic plane as a function of plane depth for XAS (blue filled curve) and XPS (red filled curve). The  (b) Left: sketch of a Rh-terminated FeRh stack close to the surface. Center: magnetic moment orientation in the AFM phase with a ferromagnetic layer as simulated in the model. Right: $\mu_{i}$ values assigned to each plane as function of its magnetic state.}
 \label{fig:fig2}
\end{figure}

In analogy with the procedure used to determine the surface termination, we developed a simple model to evaluate the thickness of the magnetic surface-layer taking advantage of the possibility of using two detection techniques with different probing depth.
In XPS experiments the probing depth of core-level photoemission can be tuned by changing the photon energy.
To optimize the magnetic signal and photon flux we used 700 eV photon energy:  the probing depth of the Fe-3p photoelectrons (at 648 eV kinetic energy)is about 1.2 nm \cite{IMFP}.
For soft X-ray absorption measured by detecting the total electron yield or the sample current, the probing depth is on the order of 3 nm \citep{Abbate1992}.
The contributions of the atomic planes to the measured XAS and XPS signal are depicted in the left and right panel of  Fig. \ref{fig:fig2}(a) respectively.
In the center of the figure the FeRh stack is reported for reference.
Since FeRh has a lattice parameter of 2.99 {\AA} the interplane distance is 1.5 \AA, hence a 50 nm thick film consists of more than 300 atomic planes.
However, only the topmost layers contribute significantly to the signal.
For XPS the first two planes contribute to the total signal for almost 22\%, while  the contribution from deeper planes becomes rapidly negligible. In the case of XAS and XMCD the first plane contributes only for less than the 5\%, and a significant contribution to the total signal derives from planes as deep as 60.
The contribution of an atomic plane to the magnetic signal is given by the product of a weight times the intensity contribution calculated as in the previous section, Eq. (\ref{eq:equa1}):
\begin{equation}
I_{a}=\sum_{i=1}^{N}I_{i,a}=\sum_{i=1}^{N}\mu_{i} \exp^{-\frac{z_{i}}{\Lambda_{a}}},
\label{eq:equa2}
\end{equation}
where $N$ is the total number of atomic planes, $\Lambda_{a}$ and $I_{a}$ are probing depth and intensity of the dichroic signal associated with a specific technique, $\mu_{i}$ is the coefficient representing the magnetic contribution, and $I_{i,a}$ is the contribution of a single plane. 
We assign a weight $\mu_{i}$ to each atomic plane. 
For ferromagnetic Fe and Rh plane $\mu_{i}$ is set to 1, while for the antiferromagnetic Fe planes and for the non-magnetic Rh planes $\mu_{i}$ is 0.
A pictorial representation of the model and of the simplest assignment of the weight values accounting for the observed magnetic signal at room temperature is given in Fig. \ref{fig:fig2}(b).
Fe and Rh intensities are calculated separately and in each case the summation index $i$ runs over the positions occupied by Fe an Rh respectively.

We calculate the intensity of the maximum magnetic signal in the FM phase separately for Fe and Rh assuming that each plane carries the maximum magnetic moment and we use this value  to evaluate the percent of signal obtained at 330 K. Using the XPS probing depth we calculate a magnetic signal equal to 35\% of the one observed in the ferromagnetic phase. This can be compared with the 31\% extracted from the experiment.  For the XAS case,  the model gives a residual moment 17.6\% for Fe and 22.1\%. for Rh to be compared with the experimental values of  18\% and 23\% respectively.
The agreement both in the case of XPS and XMCD signal with the results of the model is good and within a 10\% of uncertainty only.

\section{Computational details}

In the DFT calculations we adopted a supercell approach  (10 {\AA} vacuum was enough to prevent interactions between different replicas) using the Abinit plane-wave code \cite{abinit,Bottin2008}.
We employed norm-conserving Hartwigsen-Goedecker-Hutter (HGH) pseudopotentials \cite{Hartwigsen1998}.
Calculations were converged with a 65 Hartree cutoff (115 Hartree when semicore Fe $3s$ and $3p$ and Rh $4s$ and $4p$  were explicitly considered as valence electrons) and a 8$\times$8$\times$1 k-point grid. 
All atomic positions were relaxed until interatomic forces were less than $2.5\times10^{-3}$ eV/\AA. 
Results for the bulk are in line with previous results from literature\cite{Moruzzi1992,Lounis2003,Gruner,Gu2005,Sandratskii2011,Cherifi2014,Kudrnovsky2015} and in good agreement with experiment.
At 0 K the AFM phase is more stable by 36 meV/atom than the FM phase and the calculated lattice parameters are underestimated by 1\% with respect to experimental data \cite{zakharovFeRh}.
The magnetic moments \footnote{Obtained from the integration of the magnetisation density over a sphere of 2.1 a.u radius for Fe and 2.4 a.u. for Rh.} are 3.2 $\mu_B$ for Fe and 1.0 $\mu_B$ for Rh in the FM phase and become respectively 3.1 $\mu_B$ and 0 $\mu_B$ in the AFM phase.

\begin{figure}
  \vskip -0.0cm
    \begin{center}
          \includegraphics[width=\linewidth]{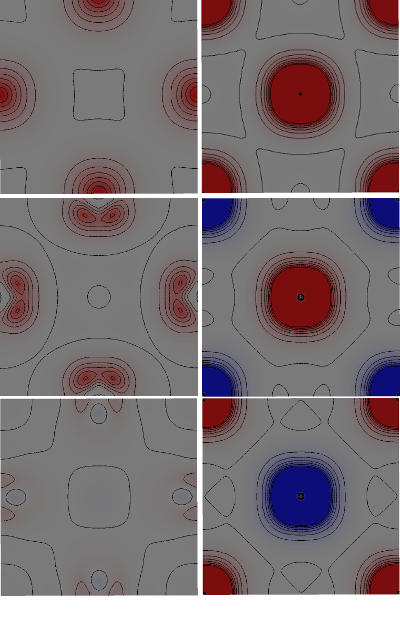}
    \end{center}
    \vskip -0.5cm
 \caption{Magnetisation density (depicted as red/blue contours) for Rh planes (panels in the left column) and Fe planes (panels in the right column).}
 \label{fig:fig6}
\end{figure} 
\section{Magnetisation density distribution}

Fig. \ref{fig:fig6} displays different sections parallel to the FeRh(100) surface of the magnetization density distribution that is reported in the bottom panel of Fig. 3 in the main text. In particular, each section corresponds to a different atomic plane of FeRh that is stacked as shown in Fig. \ref{fig:fig2}. The magnetisation density is always distributed around the atoms (not shown explicitly in the picture for sake of clarity).
The left column of the figure shows three representative planes containing Rh atoms, while the right column shows Fe planes. Concerning the Rh planes (left column), the topmost panel is the ferromagnetic Rh surface plane.
The central panel is the Rh plane that is sandwiched between a ferromagnetic and an antiferromagnetic Fe plane where Rh atoms have a magnetic moment that is half of the surface ferromagnetic Rh plane (see main text). The bottom panel is the plane where the Rh atoms have a magnetic moment equal to zero. It shows that also in this case the magnetisation is not strictly zero everywhere.
Concerning Fe planes (right column), the topmost panel is the surface ferromagnetic plane, whereas the other two panels show the alternating antiferromagnetic Fe planes in the bulk, according to the type-II antiferromagnetic configuration of FeRh.

\bibliographystyle{unsrt}
\bibliographystyle{apsrev} 

\bibliography{./FeRh_Surface_Mag_biblio_supmat}